\documentclass{article}
\usepackage[hidelinks]{hyperref} 

\usepackage[group-separator={,}]{siunitx}
\usepackage{xspace,graphicx,url,float,natbib}
\usepackage{adjustbox}
\long\def\todo#1{

}
\long\def\comment#1{\xspace}

\begin{document}
\author{Ben Klemens\thanks{
United States Treasury Office of Tax Analysis.\hfil\break 
Thanks to
Rob Axtell,
Alejandro Beltran,
David Bridgeland,
John Eiler,
Kathryn Fair, 
Robert Gillette,
Lucas Goodman, 
Sarah Haradon,
John Kaufhold,
Kye Lippold,
Eduardo Lopez, 
Eric Tressler,
and
Andy Whitten.
Thanks also to the Internal Revenue Service's Research, Applied Analytics and
Statistics Division (RAAS) for maintenance of the data and its computing environment,
and Kye Lippold and Lucas Goodman for initial development of the data set.
\hfil\break
This research was conducted by an employee of the U. S.\ Department of the
Treasury. The findings, opinions, and conclusions expressed here are entirely those of the author
and do not necessarily reflect the views or official positions of the U. S.\ Department of the Treasury.
Any taxpayer data used in this research was kept in a secured IRS
data repository, and all results have been reviewed to ensure that no confidential information is disclosed.
The data set is restricted to use on RAAS systems; contact RAAS for details of access.
}}

\date{}
\title{Measures of the Capital Network of the U.S.\ Economy}

\newcommand{\aspct}[1]{%
    \SI[round-mode=places, round-precision=1]{#1}{\percent}
}
\newcommand{\roundoff}[1]{\SI[round-mode=places, round-precision=2]{#1}\xspace} 
\newcommand{\roundoffint}[1]{\SI[round-mode=places, round-precision=0]{#1}\xspace}

\maketitle
\begin{abstract}
About two million U.S.\ corporations and partnerships are linked to each other and 
human investors by about 15 million owner--subsidiary links. %
Comparable social networks such as corporate board memberships and
socially-built systems such as the network of Internet links are ``small
worlds,'' meaning a network with a small diameter and link densities with a power-law distribution,
%
but these properties had not yet been measured for the business entity network.
%
This article shows that both inbound links and outbound links display a
power-law distribution with a coefficient of concentration estimable to
within a generally narrow confidence interval, overall, for subnetworks including only business
entities, only for the great connected component of the network, and in
subnetworks with edges associated with certain industries, for all years 2009--2021.
%
In contrast to other networks with power-law distributed link densities, 
the network is mostly a tree, and has a diameter an order of magnitude larger
than a small-world network with the same link distribution.  
%
The regularity of the power-law distribution indicates that its coefficient can
be used as a new, well-defined macroeconomic metric for the concentration of capital flows in an
economy.
%
Economists might use it as a new measure
of market concentration which is more comprehensive than measures based only on the few biggest firms.
Comparing capital link concentrations across countries would facilitate modeling the relationship
between business network characteristics and other macroeconomic indicators.\\%
\end{abstract}

Consider a successful corporation expanding into a new income stream.
It is important that liability from the new stream not impact the old, so the
new should be segregated into its own distinct legal entity.
But capital and profit and loss (P/L) may need to flow between the old entity
and the new, which is facilitated by making the old entity a partial owner of the new.
Almost three decades after the mid-1990s wave of 
state-level limited liability corporation laws \citep{Fox2005}, and the 1996
implementation of the ``check-the-box''
regulation that allowed entities to self-declare their type for tax purposes,
the economy has increasingly become a modular network of such
entities, with well-defined capital and P/L flows between them.

    In 2018, about 37,000 C corporations filed consolidated tax returns on
behalf of themselves and about 176,000 subsidiaries.
Each subsidiary is typically its own revenue stream, with capital and P/L
flows like any stand-alone business, and treating it as such provides a more appropriate
and consistent unit of analysis for questions of how capital moves through an
economy.
The entity network is an ideal vehicle for understanding the process by which
an economy builds itself, a record of both the social connections between businesspersons who
make the deals and the operational considerations of entities expanding their
lines of business.

This is the first article to describe the full network of ownership and P/L
flows from the perspective of network topology.
To date, characteristics of the full network have only been described in working papers with an eye toward tax policy
rather than global characteristics of the network
\citep{Cooper:partnerships,black:spiderweb}.
This article shows that the network of entities forming the U.S.\ economy 
has a clear power-law distribution of links, as is typical of scale-free networks.
At the micro-scale, unlike common examples of power-law networks like social
ties or Internet links, the network is close to (but not exactly) a tree, in
the sense that there is at most one path between (almost) any two nodes.

In 2017, the Tax Cuts and Jobs Act rewrote large parts of the tax code and
changed a great many of the details of the incentives behind business structuring choices,
but the fit to the power-law model held fast and the coefficient of
link concentration held relatively constant.
With one salient exception discussed in Section \ref{subindustrysec}, the same holds for graphs focusing on subindustries
within the full economy.

Using the power-law exponent as a measure of the concentration of capital
links gives us a novel and useful tool for measuring an economy.
How the concentration differs across contexts, such as different
subsectors or in economies at different levels of development, may provide
clues about what makes a growing, equitable, productive, or stable economy.

Market concentration is often measured by considering only the biggest
players in an industry (concentration ratios), or using metrics which
effectively ignore smaller values in a power-law or exponential distribution
(Hirfendahl-Hirschman index), but there is benefit to a concentration measure 
that accommodates the full range of enterprises from small to large.
For example, although there are tens of thousands of health care providers across the
U.S., the likelihood that there is
only one option for a given specialization in a given geographic area rises with industry concentration.

For tax policy and administration, examples exist where exceptionally complicated structures have been used to evade
audit, but finding ``abnormal'' and abusive arrangements requires an
understanding of what ``normal'' arrangements look like. This
article reports findings from a project aimed in that direction.

For policy considerations, the macro-level results show that the
overall distribution of links follows what some call a ``law of nature,''
\citep{FoxKeller2005}, which may be impossible to substantially change with
interventions. But the concentration of links can change over time and
may be amenable to policy interventions, should there be policy motivations to push for
more or less concentrated flows of capital.

Previous efforts at looking at, or even defining, the capital network of a domestic
economy have focused on either financial flows only among banks and
other finance-focused entities \citep{Boss2004}; or the largest (typically publicly-traded)
corporations and their boards of directors, what
are sometimes described by the authors studying them as ``élites'' \citep{Battiston2004,Corrado2006,Robins2004}.
But the economy is far more than the largest thousand firms and their owners,
and using the
\num{12051041} legal entities and human owners
with at least one connection in the 2021 entity
network of \num{15425953} links offer a more comprehensive picture.

The picture is also at a finer scale, at the level of discrete capital stocks or
lines of business, such as one entity for each restaurant location in a chain or one
subsidiary of a corporate conglomerate.

One might suspect that the number of owners or subsidiaries is simply another measure of firm size.
Below, we will see that a single entity's asset or payroll size is not well
correlated to the number of subsidiaries it has, and that assets and wages
across all entities do not conform well to a power law.
\citet{Axtell2001} found that the size of publicly-traded C corporations
does follow a power law; this will be re-evaluated in the context of the larger data
set here in Section \ref{correlationsec}.
\citet{guerrero_employment_2013} drew a network where nodes are Finnish firms
and edges are workers changing jobs from one firm to another, and found evidence
that the network has a power-law distributed node distribution and a weakly small-world exponent (in the notation
of Equation \ref{powereqn}
below, $\gamma=3.19$).
The work here provides a capital-focused complement to that result about the
labor network.

\section{ Data}
Statistics regarding the networks for tax years 2009
to 2021 will be presented below, but the exposition will primarily use the 
2021 network.

The data is from tax returns filed by C corporations, S
corporations, real estate investment trusts, and partnerships, covering the
full population of such entities.
For partnerships, all ownership is reported, but for C corporations links are
required only if the ownership share is 20\% or more, including
publicly-traded shares in C corporations.\footnote{In about 1.7\% of
      C corp observations, ownership shares below 20\% are voluntarily
      reported.}
The data set is gathered from any report of ownership in either direction on
Form 851;
Schedule B of Forms 1065 and 1120S;
Schedule K of Form 1120;
Schedule G of Form 1120; or
Schedule K-1 of Forms 1065, 1120, and 1120S.
There is redundancy, as Form 851 and Schedules B and K ask parents who their children
are, and Schedules G and K-1 ask children who their parents are.

Although business entities can own each other in almost any combination,
people are always top nodes in the network, and it is reasonable to expect that an individual makes
decisions differently from how decisions are made in a legal business entity.
Therefore, the main body of this article covers only the network with links
between U.S.\ domestic business entities (also excluding trusts, estates,
nonprofits, and links to overseas entities), which in 2021 included
\num{2230248} nodes and
\num{3925850} edges.
People and other non-businesses will be reinserted as a robustness check below.

If a partnership has income not from other partnerships, estates, or trusts;
or shows rental income, or deductions from business activity, it is classed
here as a trade or business (TB).
Non-TB operations typically make money via passive income such as interest or
rents from loaning capital or equipment to other entities, and it is a
common form to split a single business enterprise into an
active TB segment which borrows capital from a passive non-TB segment.
In 2021, among all edges between domestic business entities, \aspct{24.26} of edges are between a TB and
non-TB partnership, with a roughly even number of links where the TB
partnership is parent and where the non-TB partnership is parent.  

C corporations are taxed as distinct entities, while partnerships and
S corporations pass through P/L without a taxation step at the partnership/S
corp level. This makes C corps a disfavored choice as a node in the network:
\aspct{049.5580}
of nodes with edges in the 2021 network are TB partnerships,
\aspct{026.1742}
are non-TB partnerships,
\aspct{015.4753}
are S corporations,
and only
\aspct{008.4812} are C corporations.

\paragraph{Limitations}
Non-U.S.\ entities may be nodes at the periphery of the network,
but all links between two foreign nodes are missing from the data.

This data set excludes the large volume of capitalization via loans and
not-substantial ($<20\%$ ownership) public stock market purchases.
This is perhaps beneficial, as the remaining links are more focused on
investors closely involved in day-to-day business or (literally) heavily
invested in the operation.
From the perspective of a social network, loans and small purchases may not
reflect close social ties.

Summing across the various forms, the total ownership reported need not add to a
hundred percent(!), because of the many definitions of percent ownership,
including voting shares, and drift between so-called inside basis and outside basis.
In 2021, excluding C corporations, about \aspct{12.83} of
the data set is entities where more than 100\% ownership
is accounted for, and \aspct{7.44} with less than 90\% ownership.
Various tabulations of the subgroups with under-reporting showed
no consistent pattern.

Sole proprietorships are businesses with a single listed owner, and in 2015 they
comprised 71.9\% of non-farm business returns filed with the IRS, but only 3.8\% of
non-farm receipts. They are typically ``disregarded entities,'' meaning that
their P/L is treated as earned by the sole owner.
We take at face value the assertion by their owners that a disregarded entity
and its owner should be treated as a single unit of the owner's type
(typically a human), not as two nodes connected by an edge.
The remaining 96.2\% of the formal non-farm U.S.\ economy (by revenue, as of
2015) are nodes with edges in the network discussed in this article.

\section{ Global characteristics of the entity network}

This section presents results regarding the overall shape of the
distribution of link counts in the entity network.
Because one might expect that entities with many links are simply larger
firms, the distribution of link counts is compared to the distributions of
asset and payroll size, which will be shown to have a different shape.

\subsection{ Connected components}

Unless otherwise noted, all statistics are regarding
 the U.S. domestic business entity-only network of 2021. In that network, \aspct{060.7222} of the nodes
(excluding sole proprietorships) are in a giant connected component (GCC),
in which, treating links as undirected, there is a path from any node in the GCC
to any other in the GCC.

After the GCC of \num{1354255} nodes, the largest connected subgraph is 330 nodes,
with the great majority of the remainder of subgraphs consisting of five or fewer nodes.

\paragraph{Industry breakdown}
Table \ref{industrytab} gives the breakdown of nodes with edges in the graph
by North American Industry Classification System (NAICS) code.
NAICS codes are self-reported based on the filer's description of the primary activity of the entity.
Entities outside the U.S.\ are excluded, and the table presents the full
graph, the GCC only, the subgraph of business entities only,
and the GCC of the same subgraph.
Across all definitions of nodes in the graph, about 40\% or more
of nodes are in the business of renting real estate or equipment.
As the node set focuses on businesses embedded in the web of other
businesses, the relative percentages of entities in finance/insurance, general
management, mining, and health care rise.

\begin{table}
\centering
\makebox[\textwidth][c]{
\centering

\resizebox{7.0in}{!}{
\begin{tabular}{lcrrrr}
Industry & NAICS code & All & GCC & Entity only, all & Entity  GCC\\
Real Estate and Rental and Leasing &   53 &   \aspct{44.2976} &   \aspct{48.4761} &   \aspct{41.2908} &   \aspct{43.1899} \\
Finance and Insurance &                52 &   \aspct{10.286} &    \aspct{17.4033} &   \aspct{18.5795} &   \aspct{25.7125} \\
Professional, Scientific, and Technical Services &  54 &   \aspct{06.9238} &   \aspct{05.2618} &   \aspct{07.3976} &   \aspct{05.1619} \\
Retail Trade &                         44--45 &   \aspct{04.1547} &   \aspct{02.0968} &   \aspct{02.2887} &   \aspct{01.3903} \\
Construction &                         23 &   \aspct{04.0796} &   \aspct{02.6583} &   \aspct{03.1864} &   \aspct{02.2919} \\
Agriculture, Forestry, Fishing and Hunting &    11 &   \aspct{04.0635} &   \aspct{02.2899} &   \aspct{02.5291} &   \aspct{01.4578} \\
Nonclassifiable &                      99 &   \aspct{03.4004} &   \aspct{02.8694} &   \aspct{02.3209} &   \aspct{02.1046} \\
Accommodation and Food Services &      72 &   \aspct{03.3868} &   \aspct{03.2205} &   \aspct{02.9345} &   \aspct{02.5012} \\
Other Services (except Public Administration) & 81 &   \aspct{02.7904} &   \aspct{01.1859} &   \aspct{01.4703} &   \aspct{00.9199} \\
Health Care and Social Assistance &      62 &   \aspct{02.6932} &   \aspct{02.8297} &   \aspct{03.7633} &   \aspct{03.2319} \\
Management of Companies and Enterprises &55 &   \aspct{01.9568} &   \aspct{02.9339} &   \aspct{03.7885} &   \aspct{03.8492} \\
Manufacturing &                          31--33 &   \aspct{01.8928} &   \aspct{01.3367} &   \aspct{01.7361} &   \aspct{01.2481} \\
Wholesale Trade &                        42 &   \aspct{01.7521} &   \aspct{01.1937} &   \aspct{01.5807} &   \aspct{01.0133} \\
Administrative, Support, Waste Mgmt &   56 &   \aspct{01.7001} &   \aspct{01.0719} &   \aspct{01.4357} &   \aspct{00.9832} \\
Arts, Entertainment, and Recreation &    71 &   \aspct{01.5955} &   \aspct{01.1361} &   \aspct{01.1794} &   \aspct{00.8954} \\
Transportation and Warehousing &         48--49 &   \aspct{01.4746} &   \aspct{00.742} &    \aspct{00.9027} &   \aspct{00.5672} \\
Information &                            51 &   \aspct{01.1523} &   \aspct{00.9381} &   \aspct{01.1285} &   \aspct{00.9112} \\
Mining, Quarrying, and Oil \& Gas Extraction &  21 &   \aspct{00.9219} &   \aspct{01.2501} &   \aspct{01.3406} &   \aspct{01.4831} \\
Educational Services &                   61 &   \aspct{00.3906} &   \aspct{00.1679} &   \aspct{00.2127} &   \aspct{00.1202} \\
Utilities &                              22 &   \aspct{00.1911} &   \aspct{00.304} &    \aspct{00.3212} &   \aspct{00.4008} \\
Public Administration &                     92 &   $\sim$ 0 &    $\sim$ 0 &    $\sim$ 0 &    $\sim$ 0 \end{tabular}
}
}
\caption{Percent of nodes in a given industry, for various subsets of the
2021 network, with an increasing focus on well-connected businesses:
all nodes, the GCC of the all-nodes network, the business-entity-only
subnetwork, and the GCC of that subnetwork.
}
\label{industrytab}
\end{table}

\subsection{ Global shape}

\begin{figure}
\resizebox{\textwidth}{!}{
\includegraphics[width=2in]{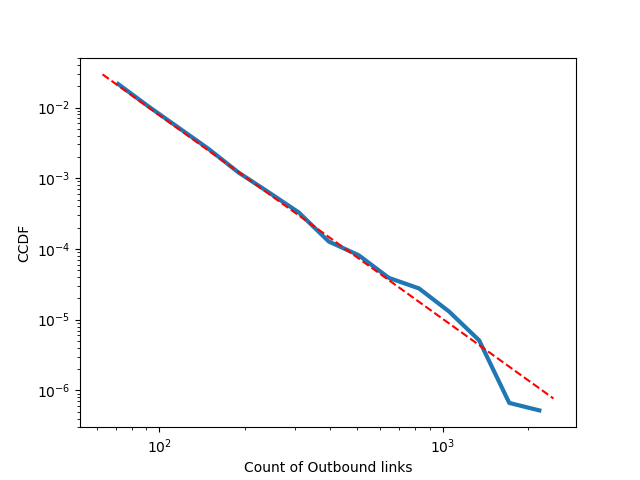}
\includegraphics[width=2in]{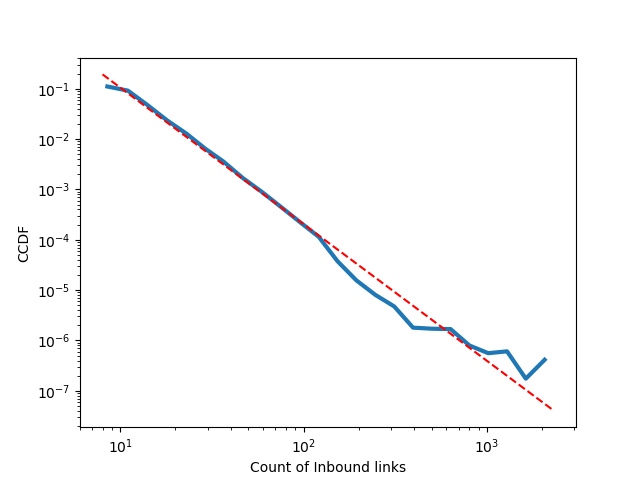}
}
\caption{Left: the distribution of outbound parent-to-child links, 2020 network, log-log scale, and its
line of best fit. Right: inbound child-from-parent links. The complementary
cumulative distribution function is the percentage of the distribution over
the given edge count. The best-fitting power law is shown as a dotted line.
} \label{linefig}
\end{figure}

The distribution of outbound link densities conforms to a power law distribution remarkably well.
Notate a power-law distribution of link counts $n$ as
\begin{equation}\label{powereqn}
\ln P(n) = -\gamma \ln(n) + K,
\end{equation}
where $P(n)$ is the likelihood that a node has $n$ links,
$K$ is a scaling constant, and $\gamma$ is a coefficient calculable from the data. 

Using the method of \citet{clauset_power}, including calculation of bootstrapped 95\%
confidence intervals,
for outbound links, $\gamma$=\roundoff{2.846587}$\pm$\roundoff{0.065482};
for inbound links,  $\gamma$=\roundoff{2.711494}$\pm$\roundoff{0.013595}.

Figure \ref{linefig} shows how
one minus the cumulative distribution function
(the complementary CDF, or CCDF)
has the hallmark linear shape on a log-log scale.\footnote{
Integrating the un-logged PDF of Equation \ref{powereqn} to a CDF,
and assuming $1-\gamma < 0$,
define the $CCDF\equiv \int_n^\infty P(x) = \frac{e^K}{\gamma-1}n^{1-\gamma}$.
This gives the log-log linear form $\ln(CCDF)=$\\
$(1-\gamma)\ln n +K-\ln(\gamma-1)$.
}

\citet{clauset_power} advises comparing the fit to a power law against a
Lognormal distribution, which is easily constructed by applying a sequence of
independent and identically distributed (IID) multiplicative shocks to a
population of initially identical
firms.\footnote{
The log of a product of IID draws is a sum of IID draws, which
by the Central Limit Theorem has a Normal distribution. That growth rate is
independent of current firm size is often referred to as {\em Gibrat's Law}.
}
The likelihood ratio tests comparing the Lognormal and power law models are
inconclusive, finding neither model fitting better than the other with
consistent statistical significance over years and network specifications.
That is, a power law model fits the data well, but the possibility that other models
also fit well is not rejected.
Section \ref{robustnesssec} will do further robustness tests.

This article uses the notation of power laws over Lognormal distributions
because of the clear interpretation of $\gamma$ as a measure of industry
concentration.
As a rule of thumb, graphs with $\gamma$ between two and three are ``small
world'' networks, with a small number of hubs and a large number of nodes with
only one or two links. The category of graphs gets its name from the short
number of steps needed to hop between any two nodes—but see below regarding
the surprisingly large diameter of the entity network.
Graphs with $\gamma\leq 2$
are extremely densely packed around the hubs, and in the other direction, as $\gamma$ grows to three and
above, the graph becomes increasingly like a diffuse random network with no 
small world properties \citep{Barabasi2016}.

\subsection{ Subindustries}
\label{subindustrysec}

For a given NAICS code, we can build a subnetwork from the set of edges
where either the parent or child has the given code.
This will be considered the network for that industry, though it will
include many entities in other fields.
For example, the health care and social assistance subnetwork 
has \aspct{37.8358} of its nodes in health care/social assistance in
2018, and \aspct{13.478} in finance.
In 2021, the portion of the network in health care/social assistance was almost identical at
\aspct{38.15328}, but the portion of the network in the finance industry
expanded by just over a quarter, to \aspct{16.9727} of the network.

\begin{figure}
\begin{adjustbox}{width=1.1\textwidth,center}
\includegraphics[width=5in]{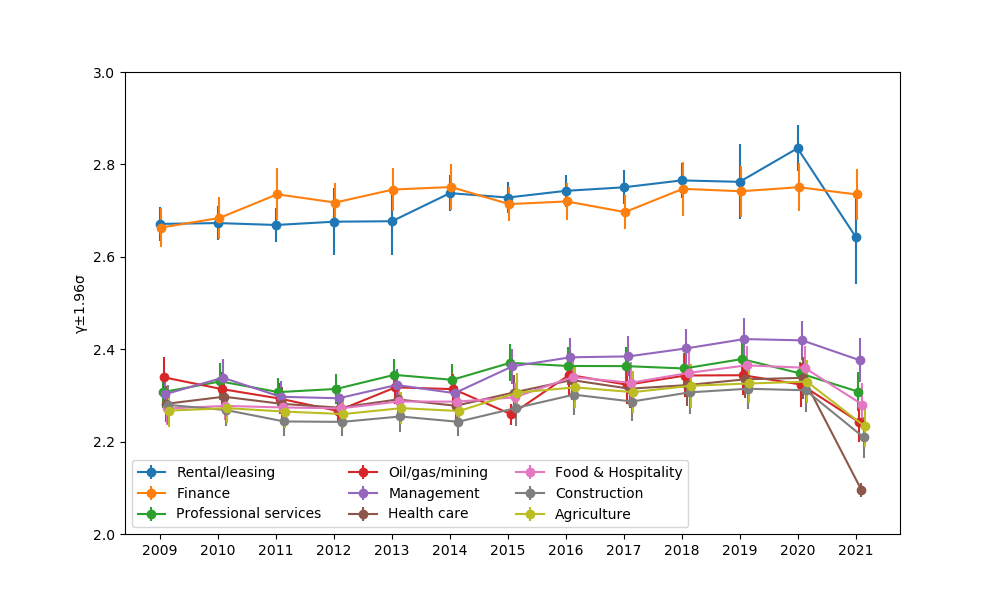}
\end{adjustbox}

\caption{Power law coefficients across time and industry subclasses.
A lower value of $\gamma$ indicates a more concentrated distribution of
capital links.
Entities classing themselves in
Finance and rental of equipment/real estate appear across the economy, but
specialized fields have values of $\gamma$ which are lower and generally
consistent.
Values are horizontally jittered to improve visibility of error bars.
} \label{industryplotfig}
\end{figure}

Figure \ref{industryplotfig} shows the values of $\gamma$ for the nine
industries with the largest presence in the GCC of the business-entity network.

The industries neatly fall into two classes. The first, rental of equipment/real
estate and finance, are industries with a lower concentration of links by the measure here.
The great majority of businesses are located somewhere, so real
estate arms of firms being distributed across the economy is no surprise;
similarly, a
great many firms have financial situations such that
segregating financial affairs into a separate business entity makes sense.

The other industries, from accounting firms to oil and gas extraction, are less generalist, and show remarkable consistency in
values of $\gamma$, both across time and across industries.

The point estimate of $\gamma$ falls for every industry from tax year 2020 to
tax year 2021, a year into the COVID-19 pandemic,
although with statistical significance in only some cases.
The stand-out exception is health care and social assistance,
which showed a statistically and possibly policy-significant increase in
concentration of capital flows.
Beyond major consolidations at the largest end of the distribution
\citep{healthmergers}, industry statistics showed an ``extraordinary'' rise in mergers \&
acquisitions even among smaller health care practices in that year \citep{pwc:deals}.

\begin{figure}
\resizebox{\textwidth}{!}{
\includegraphics[width=2in]{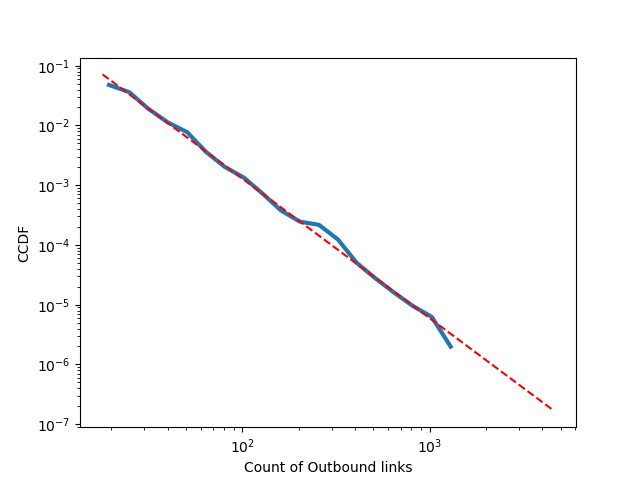}
\includegraphics[width=2in]{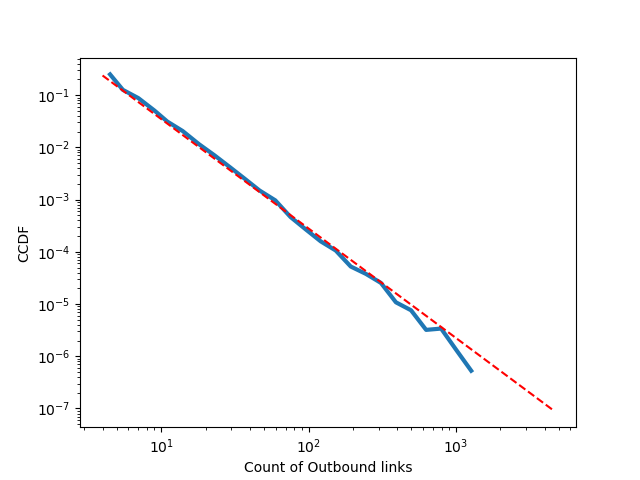}
}

\resizebox{\textwidth}{!}{
\includegraphics[width=2in]{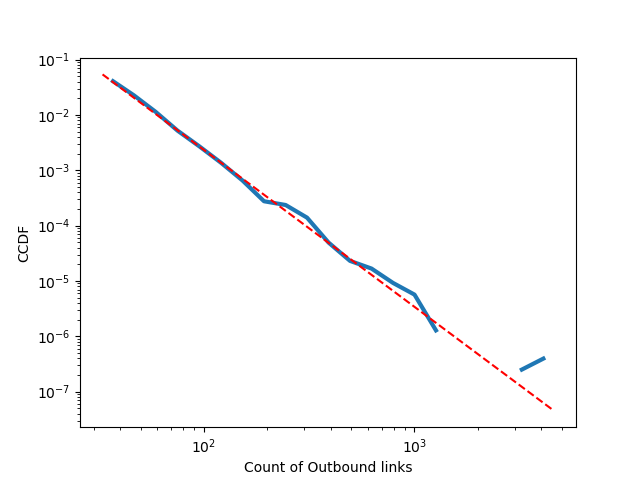}
\includegraphics[width=2in]{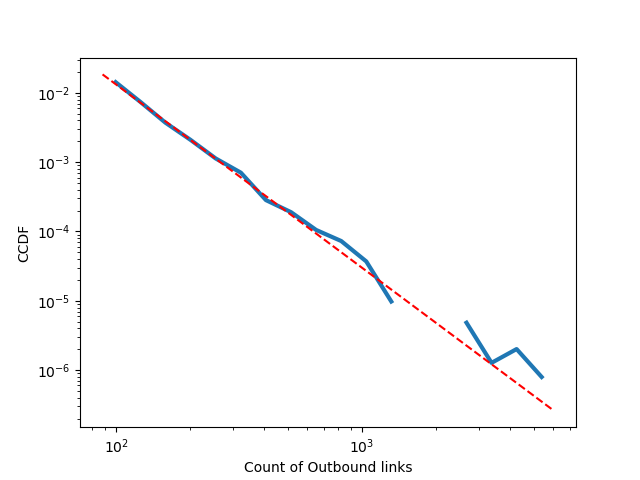}
}
\caption{Top: the CCDF of outbound links in the health care and social
services network, 2020 (left) and 2021 (right).
Bottom: the CCDF of outbound links in the real estate and equipment rental network, 2020
and 2021}
\label{healthretailplot}
\end{figure}

Other industries also showed statistically significant shifts, albeit less
dramatic ones which reversed a slow trend toward decentralization.
Figure \ref{healthretailplot} shows the CCDFs of the distribution of outbound
links and their corresponding
model best fits for health care/social services and equipment rental/real
estate, for 2020 and 2021.
The distribution of links in health care fits the straight line well from
bottom to top, though the slope of the CCDF grows steeper from 2020 to 2021,
meaning the smaller entities have relatively more of the links.
The break in the rental industry link distribution indicates a bimodal
distribution with the majority of firms being part of the
smooth distribution of links below about a thousand links, and a small
percentage of entities with many thousands of links.
That upper group grew larger in 2021, so a relatively small number of
firms may have had a disproportionate effect on the model fit and change in
$\gamma$; this is reflected in the large bootstrapped error bars in Figure
\ref{industryplotfig}.

\subsection{ Firm size}

\begin{figure}
\resizebox{\textwidth}{!}{
\includegraphics[width=2in]{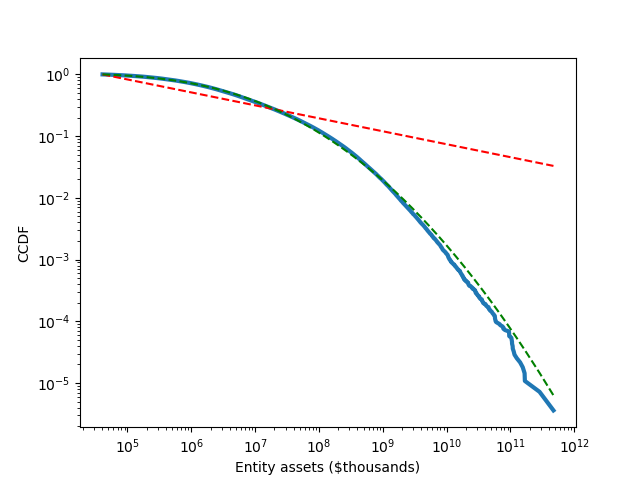}
\includegraphics[width=2in]{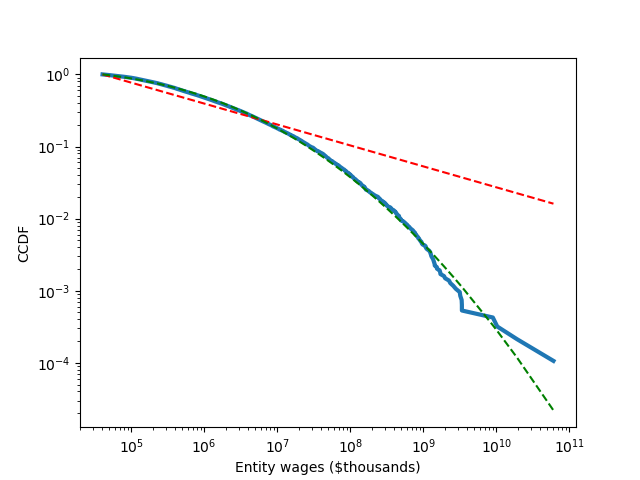}
}

\resizebox{\textwidth}{!}{
\includegraphics[width=2in]{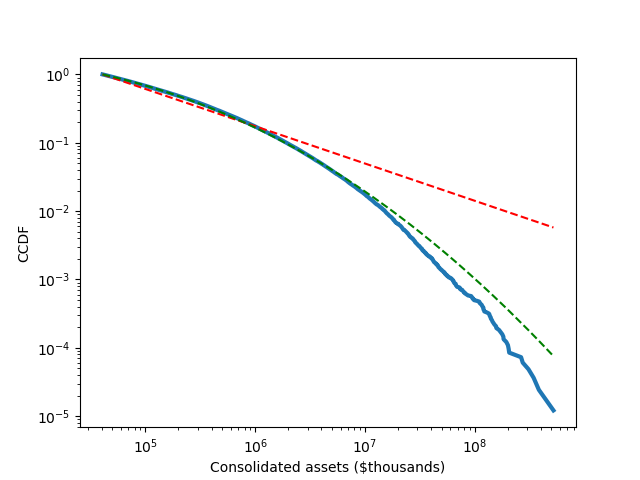}
\includegraphics[width=2in]{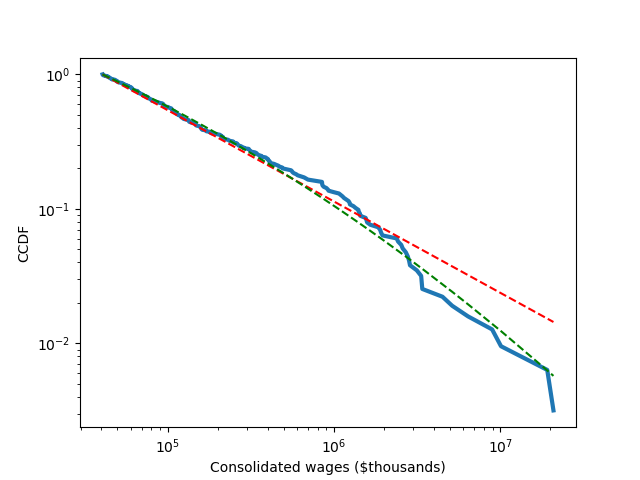}
}
\caption{Top: the CCDF of assets at left, and wages at right. Bottom
the distribution of consolidated assets at left, and consolidated wages at
right. The green curves are the best-fitting Lognormal distribution, the red
lines the best-fitting power laws.
} \label{firmsizefig}
\end{figure}

The size of firms, as measured via assets or wages, is closer to lognormal than a power law.\footnote{
      Limitations: Asset reporting (on Schedule L) is required only for firms with assets over
      \$250,000 and revenue over \$250,000.
      For 2021, \aspct{013.647} do not report any value for assets, and \aspct{054.80} report
      under \$250k in assets, including the \aspct{034.07} who report zero assets.
      There are \aspct{007.00} of entities with no asset report and gross
      income under \$1,000, which might be ultra-small businesses or might be inactive entities
      which effectively exist only on paper.
      Statistics regarding assets in the discussion here ignore all entities with null asset reports.
      } %
One means of checking this is against the distribution of assets or wages per
entity; another is to follow ownership shares, assigning all subsidiaries' assets and wages
to parents in proportion to their ownership share.
Figure \ref{firmsizefig} shows the distribution of assets and wages using
both units of analysis, and the best-fitting Lognormal and power law
distributions.
In all years, for both asset measures and entity-only wages, a likelihood ratio test indicates the Lognormal distribution is
the more likely distribution with statistical significance beyond the 99.9\%
confidence level.
To arrive at a power-law distribution, the distributions of log size would need to be
leptokurtic (have ``fatter tails'' at either end of the size distribution
relative to a Normal distribution).

\label{correlationsec}
The figures do show, however, that the consolidated wage
CCDF is relatively flat, not far from a straight-line power law. \citet{Axtell2001} used a sample of employee counts based on
consolidation reports, an exercise most comparable to this figure, and found
that subset to be closer to a power law distribution.

\begin{table}
\begin{adjustbox}{width=\textwidth,center}
\begin{tabular}{lrrrrrr}
    & $d_o$ & $d_i$ & $A_e$ & $A_c$ & $W_e$ & $W_c$ \\
Outbound degree, $d_o$ & 1 &&&&&
    \\
Inbound degree, $d_i$ & \aspct{015.1209} & 1 &&&&
    \\
Assets, $A_e$& \aspct{025.0605} & \aspct{020.5326} & 1&&&
    \\
Consolidated Assets, $A_c$ & \aspct{042.4093} &
\aspct{027.8838} & \aspct{071.1631} & 1 &&
    \\

Wages, $W_e$ & \aspct{026.4937} &
\aspct{-004.0521} & \aspct{020.4882} & \aspct{011.7702} & 1 &
    \\

Consolidated Wages, $W_c$ & \aspct{026.5321} &
\aspct{021.8068} & \aspct{025.9714} &
\aspct{039.4395} & \aspct{075.5291} & 1

\end{tabular}
\end{adjustbox}
\caption{Matrix of correlation coefficients for several measures of firm size.}
\label{confusiontab}
\end{table}

Table \ref{confusiontab} presents the correlation matrix between six
measures of size: log inbound link count, log outbound link count, log
assets, and log wages, where the last two are measured first on the entity level,
then at the consolidation level, as above.
It shows that these three concepts of firm size (links, assets, wages) are
loosely correlated, around 25\% for most measures.
Partially mechanically, the correlation between a firm's child count and
consolidated asset/payroll size is higher than the correlation between 
a firm's child count and its own size.

The measure of count of parents has some correlation with outbound degree,
assets, consolidated assets, and consolidated wages
(\aspct{015.1209}, \aspct{020.5326}, \aspct{021.8068},
and \aspct{027.8838}
correlations), but knowing how many parents an entity has tells us almost nothing about
that single entity's payroll.

\section{ Local shape}

The network is mostly a directed acyclic graph (DAG), and mostly a tree.

\subsection{ Motifs}

The motivic analysis finds almost every
type of subgraph in the capital links network, from long chains to cycles to
complete small subgraphs,
or what \citet{ShenOrr2002} refer to as dense regions of combinatorial interactions (DORs).
It is difficult to economically rationalize entities that eventually own
themselves, but cycles do exist in the entity network, with \roundoffint{2110} of
them in 2021 (rising from \num{667} in 2009).
But the great bulk of entity ownership
patterns are quotidian groups of several owners of a single child, or several children of a single parent.

\begin{figure}
\begin{adjustbox}{width=0.8\textwidth,center}
\includegraphics[width=5in]{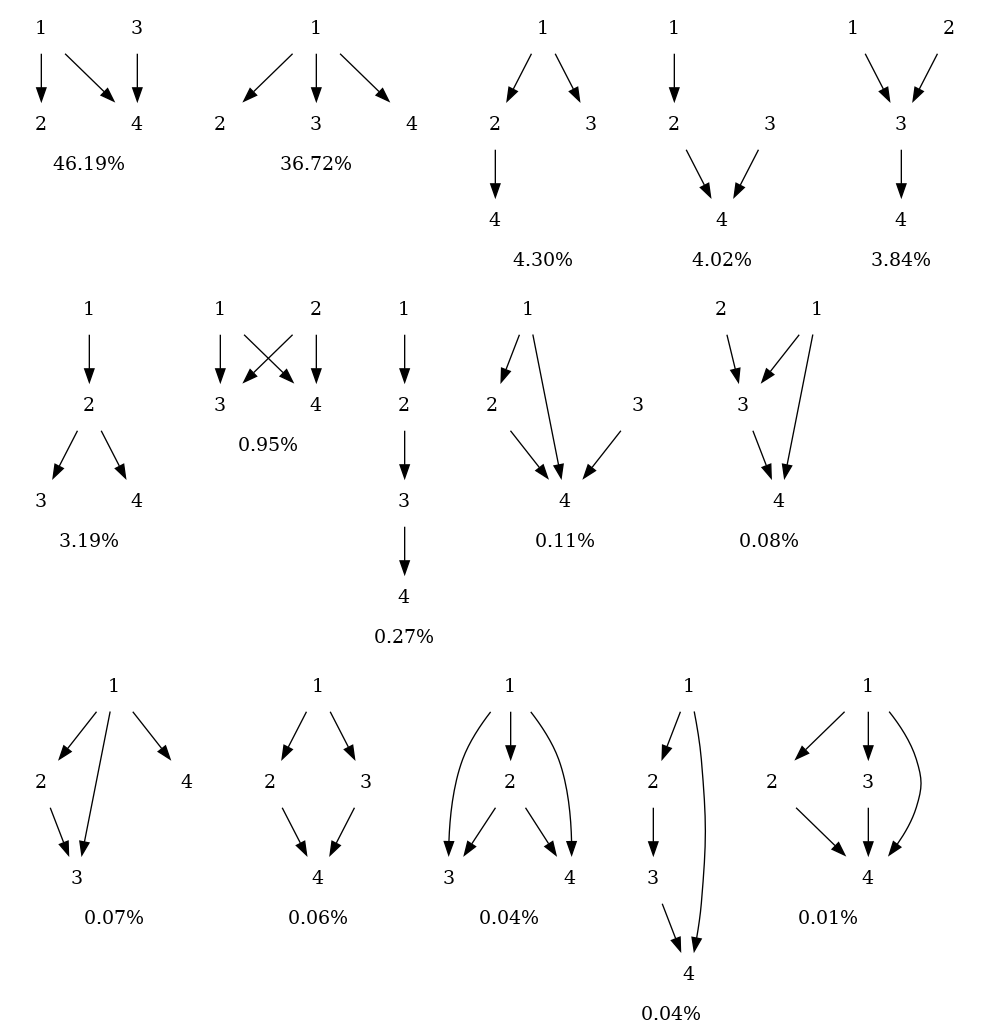}
\end{adjustbox}

\caption{The 4-node subgraphs prevalent in the giant connected component.
Frequencies are after the 71.9\% of subgraphs with
three parents and one child are excluded.
} \label{motiffourfig}
\end{figure}

A triadic census would count only
four types of trigraph in a DAG: for the 2020 network,
66.3\% two parents and one child (in the notation of
\citet{davis1967}, the 021U trigraph),
28.8\% two children and one parent (021D),
4.8\% straight-line ownership $A\to B\to C$ (021C),
and
0.13\% the directed triangle $A\to B\to C$ plus $A\to C$ (030T).

Among subgraphs of four nodes, 71.9\% are three parents and one
child, and among 5-subgraphs 71.2\% are four parents and one child, but
this is something of a combinatorial anomaly.\footnote{
A trigraph census gives equal weight to every trigraph, but this means giving
great weight to certain nodes, and this can be a problem for applications
where each node should have roughly equal consideration.

Consider a graph consisting of twenty parents $A_1\dots A_{20}$ of one child
$B$, who has one grandchild $C$. From this, one can form $(20\times 19)/2 = 190$ trigraphs
of the form $(A_i\to B, A_j\to B)$.
But there are only twenty straight-line $A_i\to B\to C$ trigraphs.
In a small-world network where there are known to be some nodes with
thousands of edges, combinations among those nodes and their adjacents
can overwhelm the trigraph census, effectively giving nearly all weight to
hub nodes.
}
Setting aside the simple funnels of one entity owned by multiple parents,
Figure \ref{motiffourfig} shows the remaining 4-subgraphs.
All possible connected DAGs without multiple paths of ownership are
displayed, all of which have likelihood greater than 0.95\%, with the exception of
the straight line $A\to B\to C\to D$ motif (whose paucity is also a
combinatorial anomaly).
Following these are all other 4-subgraphs that appear in the data, all with
likelihood 0.11\% or below, all of which have multiple paths of ownership
between a parent and child.
Among 5-node subgraphs, none with multiple chains of ownership appear with greater than 0.2\% frequency.  
Trend-wise, multiple paths of ownership (of any length) have a relatively small
count but have steadily increased over
time. The ratio of the count of
multiple-path ownerships to the count of edges (not paths, which is an exponentially larger count)
expanded
from \aspct{000.809} in the entity network of 2009
to \aspct{001.7287} in 2021.

There is nothing illegal about multiple paths of ownership, and an entity or
person with a specific interest in a sub-sub-subsidiary would find it easier
to form a second direct link than rewrite the sequence of contracts and agreements
along the original path.
Nonetheless, these indirect paths are empirically infrequent, making the
network close to a tree.
Within the GCC, the average clustering coefficient is only
\aspct{005.0414}, also reflecting the
near-tree nature of the graph.

\subsection{ Assortativity}

Among the very simple structures that form the great bulk of the
entity graph, one can easily find thousands of subgraphs wherein a handful of entities
mutually own each other in complicated arrangements.
But partnerships with large ownership counts are not isolated into certain
neighborhoods. The network is only slightly assortative:
the correlation coefficient between link counts on either side of
all edges in the 2021 entity-only network is $\rho=$\roundoff{0.029518},
steadily falling over the study period, from \roundoff{0.074590} in 2009.
\citet{Newman2002} found that socially-driven networks
like corporate director boards ($\rho=0.276$) and academic coauthors ($\rho\in[0.12,0.36]$)
generally have positive assortativity,
while mechanically driven networks like protein interactions ($\rho=-0.19$)
and the marine food web ($\rho=-0.25$) have negative assortativity.
By this measure, the financial network falls between the two regimes,
an apt position for a network built via social ties to execute the mechanics
of business operations.

\subsection{ Diameter}
For every two nodes in the same connected subnetwork, there exists a shortest
path connecting them; the diameter of a network is the length of the longest shortest path.
\citet{Bollobas2004} shows that a small-world network has
expected diameter $\ln(N)/\ln\ln(N)$, which in the case of the entity-only GCC of two
million nodes is about 5.4, not far from the ``six degrees of separation''
cliché used to describe the incredible efficiency of small-world networks.
But the diameter of the entity network varies from
32 to 
46 over the study period.
It only takes one or two firms choosing to build an exceptional ownership structure to expand the diameter,
and examples of chains several entities deep do exist in the data, possibly
deliberate efforts at obfuscation.
But
average path lengths are also high, ranging between
\roundoff{9.068394} and
\roundoff{10.855481} over the study period.
The median path length ranges between
9 and
10.

Even those many-linked hubs are only half-hubs.
The low correlation between inbound and outbound link counts,
\aspct{015.1209}, indicates that there are few network
hubs (in business terms, typically holding companies) linking a large number of inbound links to a large number of outbound
links.

Small world networks often have a small diameter because most nodes are
linked to hubs, and hubs are linked to each other.
But holding companies are not frequently buying shares of other holding companies.
The near-zero $\rho$ hints at how entities with many parents tend to be
separated in the network: a partnership with many parents may hold an entity
with one or two subsidiaries, one of which is a finance arm which has
holdings in an entity with many parents, and the pattern repeats.

The empirically observed aversion to multiple paths of ownership also
expands the diameter of the network, as there are no shortcuts between
distant nodes.

If a group of investors hold entities in a tall hierarchy, and they
acquire an interest in another set of entities in an equally tall hierarchy,
they might not expend the expense and effort to
simplify the graph, leaving a network twice as tall as the original subnetworks.

This foils any theories that the entity network has evolved for the efficient
transfer of funds across disparate entities.
Although moving information is almost always beneficial, there is rarely
reason to move a dollar from one side of the network to the other---and if the need
arises, it is easy to create a new ownership link to facilitate.

\section{ Robustness checks}
\label{robustnesssec}
Robustness of the results can be evaluated by comparing the statistic under
various situations.
Table \ref{rangestab} presents ranges for the key statistics of this article with a few
variants: including the entire graph of nodes with at
least one edge versus the GCC only; including versus excluding people;
excluding the large percentage of nodes in the finance, insurance, and
real estate (FIRE) industries; over the period from
2009–2021, covering nine tax years before the 2017 tax reform took effect in
tax year 2018, and four post-reform.

Expanding from the network of business entities to those plus their flesh-and-blood
human owners (and estates, trusts, nonprofits, and links to overseas entities)
expands the node count by about a factor of five, yet the power law
distribution continues to hold well. The power-law coefficient $\gamma$ grows
closer to that of a random graph.

At the macro scale, the patterns are consistent.
For link densities, the largest range among the eight variants is outbound
links including human and other non-business entities,
where $\max(\gamma)$ is about \aspct{12.08}
larger than $\min(\gamma)$.

The exceptionally wide diameter and median path length persists after adding
people and other non-business entities, and including or excluding out-of-GCC
groups, where the norm is only one or two links of depth.
The diameter of the network does fall by about half when FIRE-only links are excluded
from the network, indicating that such links are a key medium by which
especially long chains are formed. But the node and edge counts fall by an
order of magnitude when FIRE entities and entities whose only links are to FIRE entities are excluded.

\begin{table}
\begin{adjustbox}{width=1.2\textwidth,center}
\begin{tabular}{lrrrr}
& {\bf All, entities} & {\bf GCC, entities} & {\bf All, with people} & {\bf GCC w/o FIRE}\\

Node count
    &(\roundoffint{1749384}, \roundoffint{2230248})
    &(\roundoffint{930931}, \roundoffint{1354255})
    &(\roundoffint{8734912}, \roundoffint{12051041})
    &(\roundoffint{420148}, \roundoffint{502834})
    \\

Edge count
    &(\roundoffint{2475222}, \roundoffint{3925850})
    &(\roundoffint{1845134}, \roundoffint{3175418})
    &(\roundoffint{10763499}, \roundoffint{15425953})
    &(\roundoffint{487569}, \roundoffint{645417})
    \\
Parent link density, $\gamma$
    &(\roundoff{2.800063}, \roundoff{2.907182})
    &(\roundoff{2.784429}, \roundoff{2.895344})
    &(\roundoff{2.650937}, \roundoff{2.970944})
    &(\roundoff{2.499372}, \roundoff{2.676522})
    \\

Child link density, $\gamma$
    &(\roundoff{2.469474}, \roundoff{2.736060})
    &(\roundoff{2.456885}, \roundoff{2.719163})
    &(\roundoff{2.363150}, \roundoff{2.495179})
    &(\roundoff{2.831295}, \roundoff{3.044070})
    \\

Diameter
    &(32, 46)
    &(32, 46)
    &(31, 47)
    &(12, 21)
    \\
Median path length
    &(9, 10)
    &(9, 10)
    &(9, 12)
    &(2, 5)
    \\

Average clustering coefficient
    &(\aspct{004.2472}, \aspct{005.5702})
    &(\aspct{004.2280}, \aspct{005.5696})
    &(\aspct{002.1577}, \aspct{002.4791})
    &(\aspct{004.2280}, \aspct{005.5696})
    \\
Percent in GCC
    &(\aspct{053.2148}, \aspct{060.7222})
    &---------
    &(\aspct{038.9350}, \aspct{047.8350})
    &(\aspct{047.9587}, \aspct{051.1211})
    \\
\end{tabular}
\end{adjustbox}

\caption{The range of key statistics, 2016--2021, for the distribution of
cluster sizes excluding the GCC, link densities for outbound nodes from parents nodes, and link
densities for inbound nodes to children.} \label{rangestab}
\end{table}

\section{ Conclusion}

The entity network presented here is a representation of the full flow of capital in
the United States economy, albeit excluding very large but
network-uninformative allocations via loans and less substantial ($<20\%$ ownership) public-exchange stock purchases.
It embodies both interpersonal and operational decisions.

At the micro-scale, the graph is mostly a directed acyclic graph, and mostly
a tree, though anomalies such as
cycles and exceptionally long  paths between entities are easily found.
At the macro scale, the power-law patterns of cluster sizes and link
densities is remarkably consistent, which provides economists with an
opportunity to add a new metric to their tool kit.

This article established that the power-law coefficient $\gamma$ is a well-defined measure of
the concentration of capital ownership links in the U.S.\ economy.
What can this new measure tell us?
Figure \ref{industryplotfig} calculated $\gamma$ for 117 situations (9
industries $\times$ 13 years) and showed consistent differences for two
industries, and detected a structural shift for the health care industry one
year into the COVID-19 pandemic.
Similar comparisons could be done between $\gamma$ for U.S.\ industries
versus those in other countries at different levels of development.
A database of measures of $\gamma$ in different contexts would allow
searching for relations between $\gamma$ and other macroeconomic indicators.

Although capital and P/L flows are not stocks of wealth, and the model includes only
those with some type of capital to invest, a high concentration of flows
(low $\gamma$) may be a bellwether of deepening wealth inequality.
A high concentration indicates a network with a small number of critical
nodes; as in international capital flow networks, those nodes may be points
of failure for the network \citep{Park2021}.
Within industries, concentration measures are often used as barometers of
healthy competition and competitive pricing; such measures often focus on
only the top few competitors.
For the perfect large-$N$ theory, the ``law of nature'' of the power law tells us that both top-tier and
full-distributions should generally correlate: if there is higher
concentration at the top, there is likely higher concentration all the way
through the chain.
But the lower panels of Figure \ref{firmsizefig} gave an example where
focusing only on the top nodes in a subindustry's link distribution might tell a different story from
using $\gamma$ to directly measure and describe of the entire distribution.

This article also offers evidence for use in modeling the growth of the
business network, which falls between two different types of network
generation.
First, firms grow by amassing more assets or a
larger payroll, which as per Figure \ref{firmsizefig} follows a Lognormal
distribution, consistent with each entity seeing a central-limit-style sequence of independent
and identically distributed multiplicative shocks to its size.
Second, they may divide their organizational components into distinct entities, and
the wide diameter of the network, higher-than-expected average and median
path length, and mostly-tree nature hints at a network that has
components approximating centrally-planned corporate organization charts.

Third, they may join together entities, generating a network whose power-law distribution
of links matches that of social networks.
Partnerships and other entities are indeed {\em partnerships} between people or entities'
owners whose social ties are sufficiently close that they are willing to
share risks and capital, putting the network in the class of social,
typically small-world settings, including
outbound emails ($\gamma=2.03$, \citet{Ebel2002}), sexual networks
(for men $\gamma=2.31$, for women $\gamma=2.54$, \citet{Liljeros2001}), and
scientific collaborations ($\gamma\approx 2.5$, \citet{Newman2001}).

\bibliographystyle{plainnat}
\bibliography{motifs}
\end{document}